\documentclass[entropy,article,submit,pdftex,moreauthors]{Definitions/mdpi} 
\preto{\abstractkeywords}{\nolinenumbers}

\firstpage{1} 
\pubvolume{1}
\issuenum{1}
\articlenumber{0}
\pubyear{2024}
\copyrightyear{2024}
\datereceived{ } 
\daterevised{ } 
\dateaccepted{ } 
\datepublished{ } 
\hreflink{https://doi.org/} 

\Title{Memory corrections to  Markovian Langevin dynamics}

\TitleCitation{Memory corrections to Markovian Langevin dynamics}

\Author{Mateusz Wi{\'s}niewski $^{1}$\orcidA{}, Jerzy {\L}uczka$^{1}$\orcidB{} and Jakub Spiechowicz $^{1,}$*\orcidC{}}

\AuthorNames{Mateusz Wi{\'s}niewski, Jerzy {\L}uczka and Jakub Spiechowicz}

\AuthorCitation{Wi{\'s}niewski, M.; {\L}uczka J; Spiechowicz, J.}

\address{%
$^{1}$ \quad Institute of Physics, University of Silesia, 41-500 Chorz{\'o}w, Poland}

\corres{Correspondence: jakub.spiechowicz@us.edu.pl}

\abstract{
Analysis of non-Markovian systems and memory induced phenomena poses an everlasting challenge for physics. As a paradigmatic example we consider a classical Brownian particle of mass $M$ subjected to an external force and exposed to correlated thermal fluctuations. We show that the recently developed approach to this system, in which its non-Markovian dynamics given by the Generalized Langevin Equation is approximated by its memoryless counterpart but with the effective particle mass $M^* < M$, can be derived within the Markovian embedding technique. Using this method we calculate the first and the second order memory correction to Markovian dynamics of the Brownian particle for the memory kernel represented as the Prony series. The second one lowers the effective mass of the system further and improves precision of the approximation. Our work opens the door for the derivation of higher order memory corrections to Markovian Langevin dynamics.
}


\newcommand{\comment}[1]{}
\newcommand{\D}{\ensuremath{\mathrm{d}}}

\begin{document}

\section{Introduction} \label{sec:introduction}

Physical systems exhibiting memory are ubiquitous in nature \cite{vankampen, hanggi1995}.
The examples that attracted the interest of recent research span from quantum stochastic processes \cite{milz2020, milz2021} and quantum simulations \cite{white2020, wu2022}, to spin glasses \cite{jesi-baity2023}, active matter \cite{kanazawa2020, banerjee2022, militaru2021, narinder2018, tucci2022, cao2023}, protein folding kinetics \cite{netz}, and even animal mobility \cite{vilk2022}.
The dependence of the system present state on its past is usually consequence of its complex nature, as is the case in e.g.~viscoelastic setups \cite{ginot2022, ferrer2020, gomez2016, goychuk2012}, but the time-nonlocality may originate also from the interactions of the system with its environment and manifest as e.g.~hydrodynamic memory \cite{franosch2011, huang2010, goychuk2019}.

Dynamics of systems with memory is often modeled by non-Markovian stochastic processes and in general their complete characterization requires knowledge of an infinite set of multidimensional probability distributions.
In contrast, description of Markovian systems requires only information on their initial state and a transition probability distribution, which makes their analysis radically easier.
It is tempting to describe a complicated non-Markovian setup with its Markovian simplification, however, merely erasing the memory from the model does not allow for tracking the effects related to the non-Markovian character of the physical system. The choice is therefore between a more complete model that is generally impossible to analyze and its simplification that may not include the key features of the original setup.

It is worth mentioning two papers \cite{adelman,hanggi1978} on the problem of correspondence between a non-Markovian process with memory and its Markovian "memoryless" counterpart. Unfortunately, the exact results are obtained only for linear systems (a free Brownian particle and a harmonic oscillator which both are Gaussian processes). In these cases the original process with memory can be replaced by a nonstationary Markovian one but these two are not equivalent from the perspective of theory of stochastic processes.

Recently a new approximation method has been proposed, namely the effective mass approach \cite{Wisniewski2024-effmass}, which is a compromise between these two limiting cases. 
In this method, the original model is reduced to a Markovian one that nevertheless captures the memory effects of the original non-Markovian system.
The remnants of the memory are reflected in the effective mass of the memoryless setup.

A paradigmatic model of a system with memory is a Brownian particle exposed to correlated thermal fluctuations.
The dynamics of such a setup can be described with a Generalized Langevin Equation (GLE), in which the memory is characterized by a integral damping kernel \cite{luczka2005}.
The form of the kernel determines dissipation experienced by the particle and the correlation of thermal noise acting on it. E.g. 
In Maxwell's model of viscoelasticity, the kernel is represented as an exponentially decaying function characterized by a single relaxation time \cite{goychuk2012}.
This model can be generalized to a sum of exponentials with multiple characteristic times 
so that it can serve as an approximation for other memory profiles.
The GLE with this kernel can be represented as a multidimensional Markovian problem via the Markovian embedding procedure \cite{Marchesoni1983, Straub1986, Siegle2010a}.
In this article we develop the effective mass approach starting from the Markovian embedding technique. In doing so we derive not only the first order memory correction to the memoryless dynamics as it is done in the original work \cite{Wisniewski2024-effmass}, but also the second order one.

The paper is organized as follows. In section \ref{sec:model} we describe the model of interest, namely a Brownian particle in a correlated thermal bath.
In section \ref{sec:eff_mass} we derive the first-order correction to the approximate equation that allows to track the memory effects with a Markovian model.
Next, in Section \ref{sec:prony} we derive the second-order correction. Section \ref{sec:validation} contains their validation
Finally, we summarize the results in section \ref{sec:conclusions}.

\section{Model} \label{sec:model}

We consider a Generalized Langevin Equation describing the dynamics of a Brownian particle of  mass $M$ exposed to correlated Gaussian thermal fluctuations $\eta(t)$, which reads \cite{luczka2005}
\begin{equation} \label{eq:GLE}
    M\dot{v}(t) = - \Gamma\int_0^t K(t-s)v(s)\D s  + G(x, t) + \eta(t), 
\end{equation}
where $\Gamma$ is the friction coefficient, $K(t)$ is the memory kernel, $G(x, t)$ is an external deterministic force and the dot represents the derivative with respect to time $t$.
According to the fluctuation-dissipation theorem \cite{Kubo1966}, the kernel $K(t)$ characterizing memory of the system is related to the autocorrelation function of thermal fluctuations as 
\begin{equation} \label{eq:fde}
    \langle \eta(t) \eta(s) \rangle = \Gamma k_B T K(t-s),
\end{equation}
where $k_B$ is the Boltzmann constant and $T$ is temperature of the system. 

In the following we consider a class of integrable memory kernels $K(t)$ for which 
\begin{equation} \label{eq:norm_K}
	\int_0^\infty K(t)\mathrm{d}t = 1, \quad  \int_0^\infty t K(t)\mathrm{d}t \ \  \mbox{is finite}. 
\end{equation}
The first integral is related to the finite dissipation (damping) strength, see Equation (4.17) in Ref. \cite{Ingold}. The second one refers to the finite memory time, see Equation (4.18) in the same reference.  
The specific form of the kernel $K(t)$ depends on characteristics of environment and coupling between  thermal bath and the particle.
One of the most commonly encountered dissipation mechanisms appears in Maxwell's model of viscoelasticity, in which the particle is coupled to the environment through a spring and a dashpot connected in series \cite{goychuk2012}.
The resultant memory kernel decays exponentially and reads
\begin{equation}  \label{K}
    K(t) = \frac{1}{\tau_c}e^{-t/\tau_c},
\end{equation}
where $\tau_c$ is the memory time.
In complex environments there might be more than one time scale characterizing the relaxation of the medium.
The memory kernel can then be generalized to a sum of exponential decays, i.e. as the Prony series \cite{rheology,Baczewski2013,Mauro2018,Duong2022}
\begin{equation} \label{eq:Kexp}
    K(t) = \sum\limits_{i=1}^{N} \frac{c_i}{\tau_i} e^{-t/\tau_i},
\end{equation}
where $c_i$ are the weights of the constituents of the kernel and by virtue of the condition in Equation (\ref{eq:norm_K})
\begin{equation} \label{ci}
    \sum\limits_{i} c_i = 1.
\end{equation}
Such a memory model is known also as the Generalized Maxwell's model but it can also serve as an approximation for other, more complicated kernels \cite{goychuk2012, Siegle2010a, Siegle2010b, goychuk2019}.

\section{Effective mass approach} \label{sec:eff_mass}

The integro-differential Equation (\ref{eq:GLE}) describes the two-dimensional non-Markovian process $\{x(t), v(t)\}$. 
In Ref. \cite{Wisniewski2024-effmass} the authors show a method of approximating this equation by a much simpler memoryless Langevin one, however, with renormalized mass $M^*$ of the Brownian particle.
The derivation relies on expanding the term under the integral in Equation (\ref{eq:GLE}) into a Taylor series.
Then, if the kernel decays sufficiently fast, i.e. the memory time $\tau_c$ is short, higher order terms in $\tau_c$ can be neglected, and the integral can be approximated by two contributions -- one proportional to the velocity $v(t)$ representing the friction, and another proportional to the acceleration $\dot{v}(t)$ that introduces a correction to the particle  mass.
The resultant equation then reads \cite{Wisniewski2024-effmass}
\begin{equation} \label{eq:eff_mass}
    M^* \dot{v}(t) = -  \Gamma v(t) + G(x, t) + \xi(t),
\end{equation}
where $\xi(t)$ is the white noise obeying $\langle \xi(t)\xi(s) \rangle = 2\Gamma k_B T \delta(t-s)$, and
\begin{equation} \label{eq:dm}
    M^* = M - \Gamma\int_0^\infty tK(t)\D t = M - \Delta M
\end{equation}
is the effective mass of the particle which depends on the memory time $\tau_c$ characterizing the damping kernel $K(t)$.
The approximate Equation (\ref{eq:eff_mass}) describes a two-dimensional Markovian system $\{x(t), v(t)\}$. However,  memory effects are not completely neglected as they are represented in the mass correction $\Delta M$.

We show that for the exponentially decaying memory kernel $K(t)$ given by Equation (\ref{K}) the same result can be obtained by use of the Markovian embedding technique \cite{Marchesoni1983, Siegle2010a} which allows to convert the GLE (\ref{eq:GLE}) into a set of ordinary stochastic differential equations. Let us define the auxiliary stochastic process $w(t)$ via the relation
\begin{equation}
\label{y(t)}
w(t) = \frac{1}{\tau_c} \int_0^t \mbox{e}^{-(t-s)/\tau_c} v(s)\;ds.
\end{equation}
%
Then Equation (\ref{eq:GLE}) is transformed into the equivalent form
\begin{subequations}
\begin{align}
\label{4eq}
\dot x(t) &= v(t),  \\
M \dot v(t) &= -\Gamma w(t) + G(x, t) + \eta (t), \label{M} \\
\tau_c \dot w(t) &= -w(t) + v(t),  \label{w}\\
\label{O-U}
\tau_c \dot \eta (t) &= -\eta(t) + \xi(t),
\end{align}
\end{subequations}
where the zero-mean Gaussian white noise $\xi(t)$ obeys $\langle\xi(t)\xi(s)\rangle= 2\Gamma k_BT \; \delta(t-s)$ and the last equation of this set describes the Ornstein-Uhlenbeck noise with the exponential correlation function. 

We differentiate Equation (\ref{w}) to get  
\begin{equation} 
\tau_c \ddot w(t) = - \dot w(t) +  \dot v(t).  
\end{equation}
For $\dot w(t)$ we insert Equation (\ref{w}) yielding   
\begin{equation} 
\tau_c^2 \ddot w(t) = w(t) - v(t) + \tau_c \dot v(t).  
\end{equation}
Finally, we insert $w(t)$ into (\ref{M}) and obtain the equation
\begin{equation}
\label{equiv}
(M -  \Gamma \tau_c )\dot{v}(t) =-  \Gamma v(t) +G(x,t) + \eta(t) + \tau_c^2 \ddot w(t). 
\end{equation}
It is equivalent to the Generalized Langevin Equation (\ref{eq:GLE}). Now, we approximate it for the case of short memory time $\tau_c$. We first neglect the last term of order $\tau_c^2$. Moreover, in Equation (\ref{O-U}) we take the limit $\tau_c \to 0$ in order to preserve the correct form of the fluctuation-dissipation relation in Equation (\ref{equiv}) and consequently $\eta(t) = \xi(t)$. Then the result assumes the form (\ref{eq:eff_mass}), namely,
\begin{equation}
	M^* \dot{v}(t) = -  \Gamma v(t) + G(x, t) + \xi(t)
\end{equation}
where
\begin{equation}
	M^* = M - \Delta M = M\left(1 - \frac{\tau_c}{\tau_L}\right)
\end{equation}
is the effective mass of the particle and $\tau_L = M/\Gamma$. In the next section we show that this method can be generalized to calculate the first and second order memory corrections to the Brownian particle mass for the case of the memory kernel $K(t)$ given in the form of the Prony series.

\section{Memory kernel in the form of the Prony series} \label{sec:prony}

We extend the previous analysis to the case when the memory function $K(t)$ is represented by the Prony series, see Equation (\ref{eq:Kexp}). In this case the GLE (\ref{eq:GLE}) can be recast into an equivalent set of $N+2$ equations via the Markovian embedding scheme \cite{luczka2005}
\begin{subequations} \label{eq:set}
\begin{align}
    \dot{x}(t) &= v(t), \label{eq:set:x}\\
    M\dot{v}(t) &= -\Gamma \sum\limits_{i} w_i(t) + G(x, t) + \sum\limits_{i} \eta_i(t), \label{eq:set:v} \\
    \tau_i\dot{w}_i(t) &= -w_i(t) + c_i v(t), \label{eq:set:w}\\
    \tau_i\dot{\eta}_i(t) &= -\eta_i(t) + \xi_i(t), \label{eq:set:eta}
\end{align}
\end{subequations}
where $w_i(t)$ are auxiliary variables defined as
\begin{equation}
    w_i(t) = \frac{c_i}{\tau_i} \int_0^t e^{-(t-s)/\tau_i}v(s) \D s,
\end{equation}
and $\eta_i(t)$ are exponentially correlated Ornstein-Uhlenbeck noises
\begin{equation}
    \langle \eta_i(t) \rangle = 0,\qquad \langle \eta_i(t)\eta_j(s) \rangle = \delta_{ij} \Gamma k_B T \frac{c_i}{\tau_i}e^{-(t-s)/\tau_i}.
\end{equation}
Thus each pair $\{w_i(t), \eta_i(t)\}$ corresponds to one of the elements in the sum defining the memory kernel, c.f. Equation \ref{eq:Kexp}.
The terms $\xi_i(t)$ represent independent white noise processes obeying the relation $\langle \xi_i(t) \xi_j(s) \rangle = 2 \delta_{ij} \Gamma k_B T c_i \delta(t-s)$.
It is worth noting that the Markovian embedding is exact for a memory kernel in the form (\ref{eq:Kexp}), but it can also be applied whenever the original kernel can be approximated by a sum of exponentials.

\subsection{First order memory correction}

We follow the similar approach as in the previous section. The time derivative of Equation (\ref{eq:set:w}) gives
\begin{equation}
    \tau_i \ddot{w}_i(t) = -\dot{w}_i(t) + c_i \dot{v}_i(t).
\end{equation}
The term $\dot{w}_i(t)$ can then be eliminated by taking into account Equation (\ref{eq:set:w}) again, and therefore $w_i(t)$ can be represented as
\begin{equation}
    w_i(t) = c_i v(t) - c_i\tau_i \dot{v}(t) + \tau_i^2 \ddot{w}_i(t).
\end{equation}
If we now assume that all the memory  times $\tau_i$ are short, we can neglect the term proportional to $\tau_i^2$ and write
\begin{equation} \label{eq:1:w}
    w_i(t) \approx c_i v(t) - c_i\tau_i \dot{v}(t).
\end{equation}
Inserting it into Equation (\ref{eq:set:v}) gives
\begin{equation}
    (M - \Gamma\sum\limits_{i} c_i\tau_i)\dot{v}(t) = -  \Gamma v(t) +  G(x, t) + \sum\limits_{i} \eta_i(t), 
\end{equation}
where we used the relation (\ref{ci}). 
If we want to keep the fluctuation-dissipation relation, we have to approximate the Gaussian correlated noise by white noise, i.e. 
$\eta_i(t) \approx \xi_i(t)$, and the sum of the noise terms yields
\begin{equation} \label{eq:eta_to_xi}
    \sum\limits_{i} \eta_i(t) \approx \sum\limits_{i} \xi_i(t) = \xi(t). 
\end{equation}
The approximate Langevin equation is then the same as Equation (\ref{eq:eff_mass})
\begin{equation}
    M^*\dot{v}(t) + \Gamma v(t) = G(x, t) + \xi(t),
\end{equation}
where now the effective mass is 
\begin{equation}
    M^* = M - \Delta M_1 = M\left(1 - \sum\limits_{i} c_i \frac{\tau_i}{\tau_L}\right)
\end{equation}
and 
\begin{equation}
	\Delta M_1 = \Gamma\sum\limits_{i} c_i\tau_i
\end{equation}  
is the first order memory correction. It means that the memory effects can be reflected solely as a shift in the particle mass which is in agreement with the original result from Ref.~\cite{Wisniewski2024-effmass}.

\subsection{Second-order memory correction}

One can attempt to derive the second-order memory correction to the memoryless Langevin equation. 
It starts with taking a derivative with respect to time of Equation (\ref{eq:1:w}).
Combining the result with Equation (\ref{eq:set:v}) gives
\begin{equation}
    \dot{w}_i(t) = c_i\dot{v}(t) - \frac{c_i\tau_i}{M}\left[ -\Gamma \sum\limits_{k} \dot{w}_k(t) + \dot{G}(x, t) + \sum\limits_{k} \dot{\eta}_k(t) \right].
\end{equation}
To eliminate the time derivaties of $w_i$, $w_k$ and $\eta_k$ we utilize Equations (\ref{eq:set:w}) and (\ref{eq:set:eta}).
Then, after grouping the terms, we obtain a set of $N$ equations with $N$ auxiliary variables $w_i$
\begin{adjustwidth}{-0.5\extralength}{0cm}
\begin{equation} \label{eq:2:w}
    w_i(t) - c_i\frac{\tau_i}{\tau_L}\sum\limits_{k} \frac{\tau_i}{\tau_k} w_k = c_i\left(1 - \frac{\tau_i}{\tau_L}\sum\limits_{k} c_k\frac{\tau_i}{\tau_k}\right)v(t) - c_i\tau_i\dot{v}(t) + \frac{c_i\tau_i}{M}\sum\limits_{k} \frac{\tau_i}{\tau_k}\left(-\eta_k(t) + \xi_k(t)\right),
\end{equation}
\end{adjustwidth}
where $\tau_L = M/\Gamma$ is the Langevin time and the term involving the time derivative of the external force $G(x,t)$ was omitted since it is proportional to $\tau_i^2$ (we assume that all the correlation times $\tau_i$ are much shorter than the Langevin time $\tau_L$ and are of the same order of magnitude).
This set can also be written in the matrix form as
\begin{equation} \label{eq:2:Awb}
    \left(\mathbb{I} - \mathbf{A}\right)\mathbf{w} = \mathbf{b},
\end{equation}
where $\mathbb{I}$ is the identity matrix and 
\begin{equation}
    \mathbf{A}_{ij} = a_i\frac{\tau_i}{\tau_j},\qquad a_i = c_i\frac{\tau_i}{\tau_L}. 
\end{equation}
Moreover, $\mathbf{w}$ is a vector of the auxiliary variables $w_i$  and $\mathbf{b}$ consists of the right-hand side terms of the set (\ref{eq:2:w}).
The inverse of the matrix $\mathbb{I} - \mathbf{A}$ reads
\begin{equation}
    \left[(\mathbb{I} - \mathbf{A})^{-1}\right]_{ij} = \frac{1}{1-\sum\limits_{k} a_k}
    \begin{cases}
        1 - \sum\limits_{k\neq i} a_k &\text{if } i = j,\\
        \mathbf{A}_{ij} &\text{if } i \neq j
    \end{cases}
\end{equation}
Equation (\ref{eq:2:Awb}) can then be solved by calculating $(\mathbb{I}-\mathbf{A})^{-1}\mathbf{b}$, and the auxiliary variables $w_i(t)$ read
\begin{equation}
    w_i(t) = c_i v(t) - c_i\tau_i\left(1 + \frac{\tau_i/\tau_L}{1-\sum\limits_{k} a_k}\right)\dot{v}(t) + \frac{1}{1-\sum\limits_k a_k} \frac{c_i\tau_i}{M}\sum\limits_{k} \frac{\tau_i}{\tau_k}\left(-\eta_k(t)+\xi_k(t)\right).
\end{equation}
After inserting it into Equation (\ref{eq:set:v}) and rearranging the terms one gets
\begin{adjustwidth}{-0.87\extralength}{0cm}
\begin{equation}
    \left[M - \Gamma\sum\limits_i c_i\tau_i\left(1 + \frac{\tau_i/\tau_L}{1-\sum\limits_{k} a_k}\right)\right]\dot{v}(t) + \Gamma v(t)\sum\limits_i c_i \approx G(x, t) + \frac{1}{1-\sum\limits_k a_k}\sum\limits_i a_i\sum\limits_k \frac{\tau_i}{\tau_k}\left(\eta_k(t) - \xi_k(t)\right) + \sum\limits_i \eta_i(t).
\end{equation}
\end{adjustwidth}
Here the friction term again reads $\Gamma v(t)$ (see Equation (\ref{ci})), so to keep the fluctuation-dissipation relation  we must again 
use the approximation $\eta_k(t) \approx \xi_k(t)$ (see Equation (\ref{eq:eta_to_xi})).
This results in the memoryless Langevin equation
\begin{equation}
    M^*\dot{v}(t) + \Gamma v(t) = G(x, t) + \xi(t),
\end{equation}
where
\begin{equation}
    M^* = M - \Delta M_2
\end{equation}
is the effective mass of the particle and the second-order correction reads 
\begin{equation}
    \Delta M_2 = \Gamma\sum\limits_{i} c_i\tau_i\left(1+\frac{\tau_i}{\tau_L-\sum\limits_{k} c_k\tau_k}\right) = \Delta M_1 + \Gamma\sum\limits_i \frac{c_i\tau_i^2}{\tau_L - \sum\limits_k c_k\tau_k}.
\end{equation}
Since we assume that all the correlation times $\tau_k \ll \tau_L$, the mass correction can be simplified to the expression 
\begin{equation}
    \Delta M_2 = \Delta M_1 + \Gamma \sum\limits_i \frac{c_i \tau_i^2}{\tau_L}
\end{equation}
for which the effective mass reads
\begin{equation}
    M^* = M\left[1 - \sum\limits_i c_i\frac{\tau_i}{\tau_L}\left(1 + \frac{\tau_i}{\tau_L}\right)\right].
\end{equation}
It is worth noting that for $\tau_k \ll \tau_L$ the mass correction $\Delta M_2$ is always greater than $\Delta M_1$.

\section{Verification of the memory corrections} \label{sec:validation}

Now we would like to discuss validation of the above presented approach to non-Markovian dynamics.
For this purpose we consider a Brownian particle moving in the spatially periodic potential $U(x) = U_0\sin(2\pi x/L)$ and  driven by the periodic force $A\cos(\Omega t)$ as well as the static bias $F$.
The external force $G(x, t)$ thus reads \cite{entropy,wisniewski2022,wisniewski2023}
\begin{equation}
    G(x, t) = -U'(x) + A\cos(\Omega t) + F,
\end{equation}
where the prime represents the derivative with respect to the position $x$.
Moreover we choose the simplest case of the kernel $K(t)$ consisting of a single exponential decay, i.e.
\begin{equation} \label{eq:K1}
    K(t) = \frac{1}{\tau_c}e^{-t/\tau_c},
\end{equation}
which corresponds to $N=1$, $c_1=1$ and $\tau_1=\tau_c$ in Equation (\ref{eq:Kexp}). The quantity of prime interest is the asymptotic long time average velocity of the particle
\begin{equation}
	\langle v \rangle = \lim\limits_{t\to\infty} \frac{1}{t} \int_0^t \langle \dot{x}(s) \rangle \D s,
\end{equation}
where the brackets denote averaging over the initial conditions and realizations of the thermal noise \cite{spiechowicz2016scirep}.

Following \cite{Wisniewski2024-effmass} we introduce new dimensionless variables
\begin{equation}
	\hat{x} = \frac{x}{L},\quad \hat{t} = \frac{t}{\tau_0},\quad \hat{v}(\hat{t}) = \frac{\tau_0}{L}v(t),\quad \hat{w}(\hat{t}) = \frac{\tau_0}{L}w(t),\quad \hat{\eta}(\hat{t}) = \frac{L}{V_0}\eta(t),
\end{equation}
where $\tau_0 = \Gamma L^2/U_0$.
The set of Equations (\ref{eq:set}) with the kernel given by Equation (\ref{eq:K1}) can be then recast to
\begin{subequations} \label{eq:set-dimless}
\begin{align}
    \dot{\hat{x}}(\hat{t}) &= \hat{v}(t),\\
    m\dot{\hat{v}}(\hat{t}) &= -\hat{w}(\hat{t}) - \hat{U}'(\hat{x}) + a\cos(\omega \hat{t}) + f + \hat{\eta}(\hat{t}),\\
    \tau \hat{w}(\hat{t}) &= -\hat{w}(\hat{t}) - \hat{v}(\hat{t}),\\
    \tau \hat{\eta}(\hat{t}) &= -\hat{\eta}(\hat{t}) + \hat{\xi}(\hat{t}),
\end{align}
\end{subequations}
where the dimensionless parameters are
\begin{equation}
    m = \tau_L/\tau_0,\ \ \hat{U}(\hat{x}) = \sin(2\pi\hat{x}),\ \ a = AL/V_0,\ \ \omega = \Omega\tau_0, f = FL/V_0 \text{  and  } \tau = \tau_c/\tau_0.
\end{equation}
The white noise term $\hat{\xi}(\hat{t})$ obeys the relation $\langle \hat{\xi}(\hat{t}) \hat{\xi}(\hat{s}) \rangle = 2D\delta(\hat{t}-\hat{s})$, where $D = k_B T/U_0$.
The approximate equation in turn reads
\begin{equation} \label{eq:effmass-dimless}
    m^*\dot{\hat{v}}(\hat{t}) + \hat{v}(\hat{t}) = -\hat{U}'(\hat{x}) + a\cos(\omega\hat{t}) + f + \hat{\xi}(\hat{t}),
\end{equation}
where $m^* = m - \Delta m_i$ ($i = 1,2$) is the effective mass of the system and the dimensionless memory correction reads
\begin{equation} \label{eq:dm1}
    \Delta m_1 = \tau
\end{equation}
in the first order, and 
\begin{equation} \label{eq:dm2}
    \Delta m_2 = \Delta m_1 + \frac{\tau^2}{m - \tau} = \tau\left(1 + \frac{\tau}{m-\tau}\right)
\end{equation}
in the second order.
For clarity, we omit the hat over the rescaled variables and write $t$ instead of $\hat{t}$, $x$ in place of $\hat{x}$, and so on.

To compare the effective mass approach with the first- and second-order memory corrections, we chose the following same parameter set: $m = 1$, $a = 10$, $\omega = 4$, and $D = 10^{-3}$. Then we implemented a weak second-order predictor-corrector algorithm \cite{Platen} and solved numerically the set of Equations (\ref{eq:set-dimless})  and Equation (\ref{eq:effmass-dimless})  for $t\in[0,\ 5\!\times\!10^{3}\tau_{\omega}]$ with a time step $h=2.5 \times 10^{-3}\tau_{\omega}$, where $\tau_{\omega} = 2\pi/\omega$ is the period of the dimensionless driving force $a\cos(\omega t)$. The particle's velocity was then averaged over last $1000\tau_{\omega}$ time units as well as $2^{18}$ realizations of the thermal noise and initial conditions $x(0)$ and $v(0)$ distributed uniformly over the intervals $[0, 1]$ and $[-2, 2]$, respectively.
To parallelize the calculations, the simulations were performed on modern Graphics Processing Units, which shortened the computational time by several orders of magnitudes compared to the traditional approach involving Central Processing Units \cite{spiechowicz2015}.

\begin{figure}[t]
	\centering
	\includegraphics[width=0.6\linewidth]{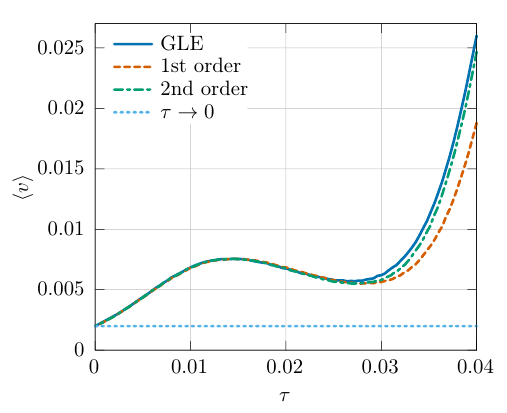}
	\caption{Average velocity $\langle v \rangle$ of the Brownian particle as a function of the memory time $\tau$ for the original GLE and the approximate equation with first- and second-order correction. The memoryless limit $\tau \to 0$ was also depicted for reference.}
	\label{fig:v_tm}
\end{figure}

Fig.~\ref{fig:v_tm} shows the average velocity $\langle v \rangle$ as a function of the correlation time $\tau$.
The solid blue curve represents the numerical solution of the original GLE (set \ref{eq:set-dimless}).
The dependence of $\langle v \rangle$ on $\tau$ is non-monotonic, but in general the value of the average velocity is greater in the presence of memory $\tau\in(0, 0.04)$ than in the memoryless limit $\tau\to0$ (light-blue dotted curve).
This means that the naive Markovian approximation obtained for $\tau\to0$ does not correctly reproduce the dynamics of the particle.

The effects of short memory can however be reproduced in the memoryless Langevin dynamics (\ref{eq:effmass-dimless}) with a properly modified particle mass.
The red dashed curve shows the solution of Equation~(\ref{eq:effmass-dimless}) for the same set of parameters and the mass correction given by Equation~(\ref{eq:dm1}).
For small values of $\tau$ the approximate curve is barely distinguishable from the original one.
When the correlation time increases, the qualitative behavior remains the same, but the accuracy becomes worse.
For $\tau > 0.025$ the mass correction seems to be too small, and for this reason the approximate curve ``lags'' behind the original one.

The quantitative precision of the approximation can be improved by taking into account the second-order correction to the particle mass.
The green dashed-dotted curve shows the $\langle v \rangle(\tau)$ function for Equation~(\ref{eq:effmass-dimless}) with the mass correction given by Equation~(\ref{eq:dm2}).
Since $\Delta m_2 > \Delta m_1$, the ``lag'' is reduced, and the approximate characteristic stays close to the original curve even for $\tau > 0.025$.
This means that the second-order correction to the particle mass extends the range of the values of the correlation time $\tau$ for which the effective mass approach correctly predicts the dynamics of the system with memory.

\section{Conclusions} \label{sec:conclusions}
Summarizing, in this work we presented new derivation of the effective mass approach to memory in non-Markovian systems that is based on the Markovian embedding technique. In doing so we considered a Brownian particle subjected to an external force and exposed to thermal fluctuations, whose autocorrelation function is given as a sum of $N$ exponential decays. Such a non-Markovian system described by the Generalized Langevin equation can be represented as multidimensional Markovian one upon introducing $N$ auxiliary variables via the Markovian embedding method. Using this representation we derived the memoryless Langevin equation in which the memory effects are reflected solely in the change of the system mass.

First, we showed that the first-order memory correction to the particle mass coincides with the result of the effective mass approach derived with a different method \cite{Wisniewski2024-effmass}.
Next, we derived a second-order memory correction that lowers the effective mass of the system.
We verified that both approximations correctly reproduce the dynamics of the original system as long as the correlation time of the fluctuations is short.
Moreover, we showed that taking into account the second-order memory correction improved the accuracy of the effective mass approximation.

The approach presented in this article provides an efficient method of studying non-Markovian dynamics that is often demanding both in analytical and numerical treatment.
The simplified representation of the Generalized Langevin Equation reduces the computational cost of the numerical analysis of the system with memory and allows for a more comprehensive exploration of its parameters space.
The new method of determining the effective mass opens the door for the derivation of higher-order memory corrections to Markovian dynamics.
\vspace{6pt} 

\authorcontributions{Conceptualization, M.W. and J.S.; methodology, M.W. J.{\L}. and J.S.; software, M.W. and J.S.; validation, M.W.; formal analysis, M.W. and J.S.; investigation, M.W.; resources, J.S.; data curation, M.W.; writing---original draft preparation, M.W.; writing---review and editing, J.{\L}. and J.S.; visualization, M.W.; supervision, J.S.; project administration, J.S.; funding acquisition, J.S. All authors have read and agreed to the published version of the manuscript.}

\funding{This research was funded by NCN grant number 2022/45/B/ST3/02619 (J.S.).}

\dataavailability{We encourage all authors of articles published in MDPI journals to share their research data. In this section, please provide details regarding where data supporting reported results can be found, including links to publicly archived datasets analyzed or generated during the study. Where no new data were created, or where data is unavailable due to privacy or ethical restrictions, a statement is still required. Suggested Data Availability Statements are available in section ``MDPI Research Data Policies'' at \url{https://www.mdpi.com/ethics}.}

\conflictsofinterest{The authors declare no conflicts of interest.} 

\begin{adjustwidth}{-\extralength}{0cm}
\reftitle{References}

\PublishersNote{}
\end{adjustwidth}
\end{document}